\documentclass{appolb}

\usepackage{graphicx}
\usepackage{latexsym} 
\usepackage{epstopdf}

\newcommand{\be}{\begin{equation}}
\newcommand{\ee}{\end{equation}}
\newcommand{\ba}{\begin{eqnarray}}
\newcommand{\ea}{\end{eqnarray}}
\newcommand{\ban}{\begin{eqnarray*}}
\newcommand{\ean}{\end{eqnarray*}}

\begin{document}

\title{Integrated Azimuthal Correlations \\ in Nucleus--Nucleus Collisions
at CERN SPS\thanks{Presented by St. Mr\'owczy\'nski at the HIC-for-FAIR Workshop 
\& XXVIII Max-Born Symposium `Three Days on Quarkyonic Island', Wroc\l aw, 
Poland, May 18-21, 2011.}
}

\author{Katarzyna Grebieszkow
\address{Faculty of Physics, Warsaw University of Technology, \\
ul. Koszykowa 75, PL - 00-662 Warsaw, Poland}
\and
Stanis\l aw Mr\'owczy\'nski
\address{Institute of Physics, Jan Kochanowski University, \\
ul. \'Swi\c etokrzyska 15, PL - 25-406 Kielce, Poland \\
and National Centre  for Nuclear Research, \\
ul. Ho\.za 69, PL - 00-681 Warsaw, Poland}
}

\date{October 21, 2011}

\maketitle
\begin{abstract}

Azimuthal correlations of particles produced in nucleus-nucleus collisions
at CERN SPS are discussed. The correlations quantified  by the integral 
measure $\Phi$ are shown to be dominated by effects of collective flow. 

\end{abstract}

\vspace{0.3cm}





\section{Introduction}

There are various sources of azimuthal correlations of particles produced in relativistic 
heavy-ion collisions. One mentions here jets and minijets resulting from (semi-)hard parton-parton 
scattering and collective flow due to the cylindrically asymmetric pressure gradients, see 
the review articles \cite{CasalderreySolana:2007zz} and \cite{Voloshin:2008dg}, respectively. 
More exotic sources of correlations are also possible. As argued in \cite{Mrowczynski:2005gw},
the plasma instabilities, which occur at an early stage of collisions, can generate
the azimuthal fluctuations. Except the dynamically interesting mechanisms, there
are also rather trivial effects caused by decays of hadronic resonances or by
energy-momentum conservation.

Several methods has been developed to study fluctuations on event-by-event basis.
In particular, the so-called measure $\Phi$ proposed in \cite{Gazdzicki:1992ri}
was used to measure the transverse momentum \cite{Anticic:2003fd,Anticic:2008vb} 
and electric charge fluctuations \cite{Alt:2004ir}. The measure proved to be very sensitive 
to dynamical correlations and it was suggested to apply it to study azimuthal ones
\cite{Mrowczynski:1999vi}.  Such an analysis is underway using experimental data accumulated 
by the NA49 and NA61 Collaborations and some preliminary results are already published
\cite{Cetner:2010vz}. 

The fact that the measure $\Phi$ is sensitive to correlations of various origin is advantage and 
disadvantage at the same time. A signal of correlations can be rather easily observed but it is 
difficult to disentangle different contributions. For this reason we studied in \cite{Cetner:2010wr} 
how various sources of azimuthal correlations contribute to the measure $\Phi$. And here we 
make use of the study \cite{Cetner:2010wr} to interpret the preliminary experimental results  
\cite{Cetner:2010vz}.  We show that the observed integrated correlations are mostly generated 
by the collective flow.

\section{Measure $\Phi$}

Let us first introduce the correlation measure $\Phi$. One defines the variable
$z\buildrel \rm def \over = x - \overline{x}$, where $x$ is a single particle's
characteristics such as the particle transverse momentum, electric charge or azimuthal 
angle. The overline denotes averaging over a single particle inclusive distribution.
In the subsequent sections, $x$ will be identified with the particle azimuthal angle
$\phi$ and the fluctuation measure will be denoted as $\Phi_\phi$. The event 
variable $Z$, which is a multiparticle analog of $z$, is defined as 
$Z \buildrel \rm def \over = \sum_{i=1}^{N}(x_i - \overline{x})$, where the summation
runs over particles from a given event. By construction, $\langle Z \rangle = 0$,
where $\langle ... \rangle$ represents averaging over events (collisions). The
measure $\Phi$ is finally defined as
\be
\label{phi-def}
\Phi \buildrel \rm def \over =
\sqrt{\langle Z^2 \rangle \over \langle N \rangle} -
\sqrt{\overline{z^2}} \;.
\ee
It is evident that $\Phi = 0$, when no inter-particle correlations are present.
The measure also possesses a less trivial property - it is {\it independent} of
the distribution of the number of particle sources if the sources are identical
and independent from each other. Thus, the measure $\Phi$ is `blind' to the
impact parameter variation as long as the `physics' does not change with the
collision centrality. In particular, $\Phi$ is independent of the impact parameter
if the nucleus-nucleus collision is a simple superposition of nucleon-nucleon
interactions. 

\begin{figure*}[t]
\centering
\includegraphics*[width=10cm]{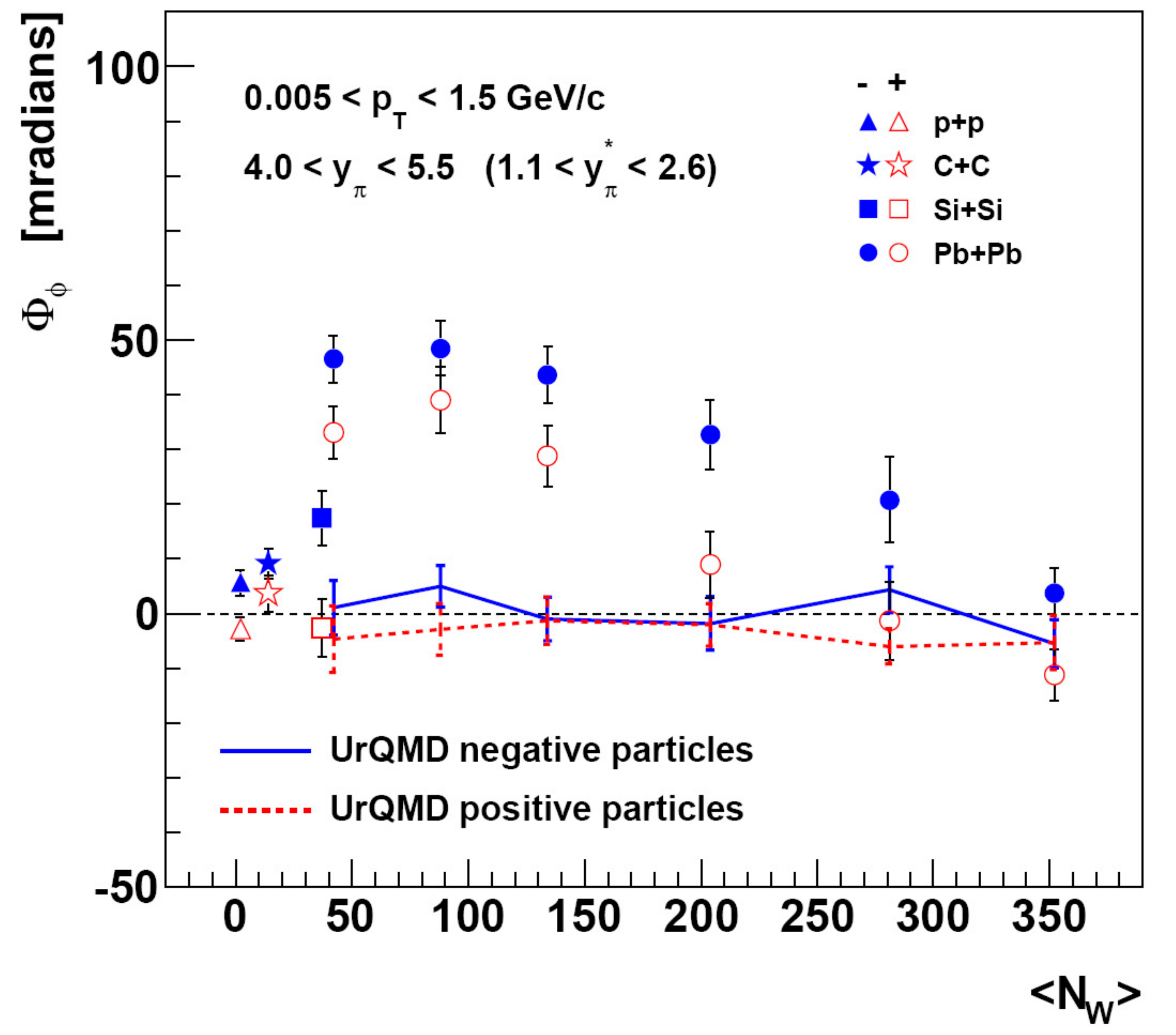}
\caption{$\Phi_\phi$ as a function of the number of wounded nucleons for 
positively and negatively charged particles produced in various colliding 
systems: p+p, C+C, Si+Si and Pb+Pb at 158A GeV. }
\label{fig-data-NW}
\end{figure*}

\section{Experimental data}

As already mentioned, the NA49 Collaboration undertook an effort to study $\Phi_\phi$
in nucleus-nucleus collisions at CERN SPS and some preliminary results are already
published \cite{Cetner:2010vz}. In Fig.~\ref{fig-data-NW} we show $\Phi_\phi$ as a function 
of the number of wounded nucleons for  positively and negatively charged particles 
produced in various colliding systems (p+p, C+C, Si+Si and Pb+Pb) at 158 GeV per nucleon
which is the top SPS energy. In the case of Pb+Pb collisions, a whole sample of events
was split into six centrality classes. The measurement was performed in a rather limited domain 
of rapidity which in the laboratory frame is $4.0 \le y_\pi \le 5.5$. It corresponds to the
center-of-mass rapidity interval $1.1 \le y_\pi^* \le 2.6$. To determine particle's rapidity the 
pion mass was assigned to all particles. The acceptance window in particle's transverse 
momentum was fairly broad ($0.005 \le p_T \le 1.5~{\rm GeV}/c$) but in azimuthal angle 
was incomplete, as in the NA49 measurement of transverse momentum fluctuations
\cite{Anticic:2003fd}.

As seen in Fig.~\ref{fig-data-NW}, significant positive values of $\Phi_\phi$ are observed with 
a maximum at $\langle N_w \rangle \approx 50$ in Pb+Pb interactions. The correlations are 
very small for both the smallest and highest numbers of wounded nucleons. One also observes 
that $\Phi_\phi$ is higher for negative particles than for positive ones. 

\begin{figure*}[t]
\centering
\includegraphics*[width=10cm]{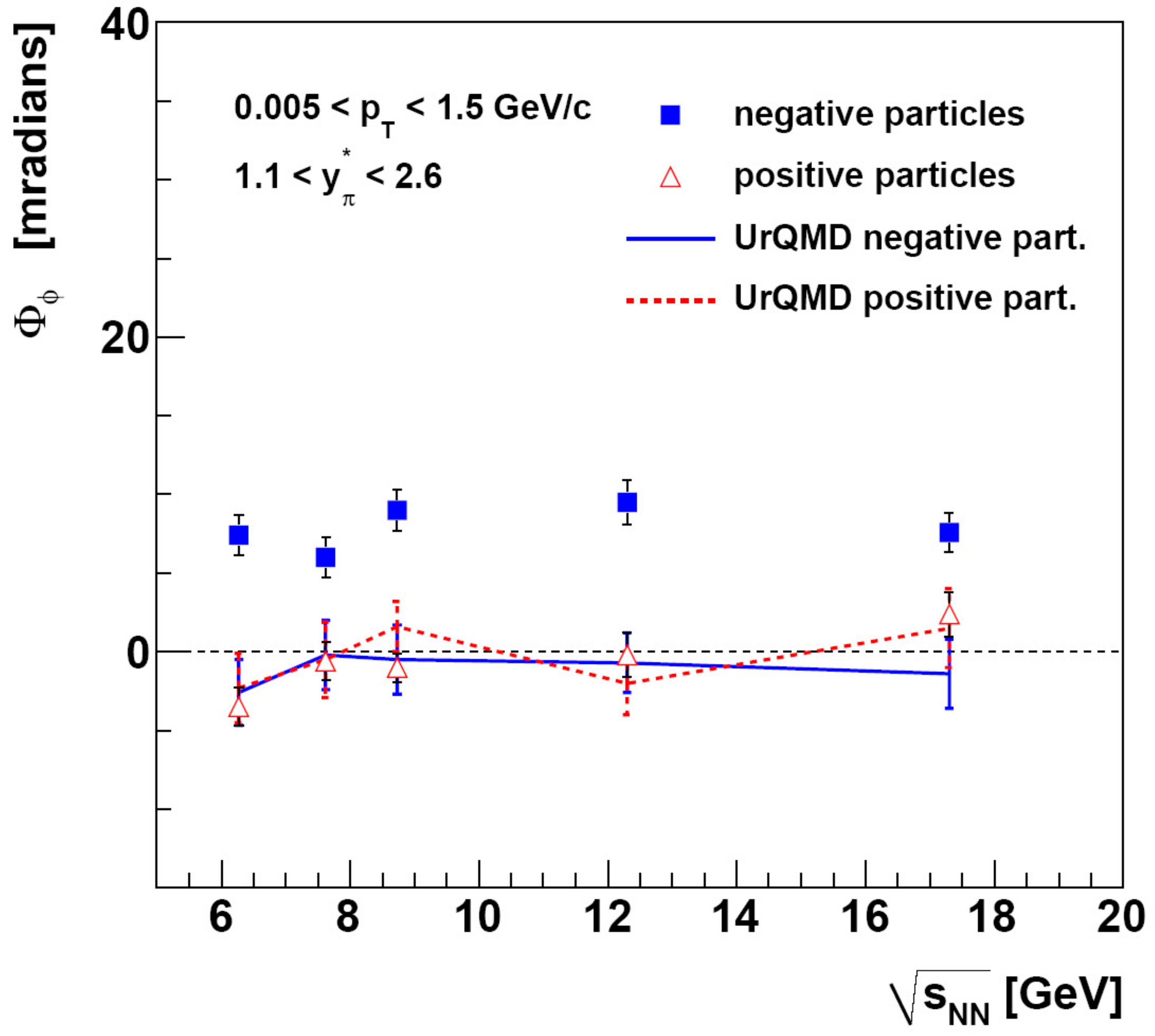}
\caption{$\Phi_\phi$ as a function of the colliding center-of-mass energy of nucleon-nucleon
system for positively and negatively charged particles produced in most central 
Pb+Pb collisions. }
\label{fig-data-energy}
\end{figure*}

Fig.~\ref{fig-data-energy} shows the energy dependence of $\Phi_\phi$ for the 7.2\% most 
central Pb+Pb interactions. The produced particles were registered in the fixed interval
of center-of-mass rapidity $1.1 \le y_\pi^* \le 2.6$ and the acceptance was as in the NA49 
measurement of collision energy dependence of transverse momentum fluctuations
\cite{Anticic:2008vb}. As seen, the values of $\Phi_\phi$ for positive particles are consistent 
with zero but for negative particles $\Phi_\phi$ is positive. No collision energy dependence 
of the correlations is observed.

In Figs. \ref{fig-data-NW} and \ref{fig-data-energy} we also show  predictions of the UrQMD 
model \cite{Bass:1998ca,Bleicher:1999xi}. Since an orientation of the reaction plane is
fixed for all collisions in the model, it was randomly rotated to compare the model predictions 
with the experimental data. As seen in Figs. \ref{fig-data-NW} and \ref{fig-data-energy},
the model  provides vanishing values of $\Phi_\phi$ within the experimental acceptance. Therefore, 
the mechanism responsible for the correlation signal is either missing or not strong enough in
the  UrQMD  model.

\section{Interpretation of experimental data}

\begin{figure*}[t]
\centering
\includegraphics*[width=10cm]{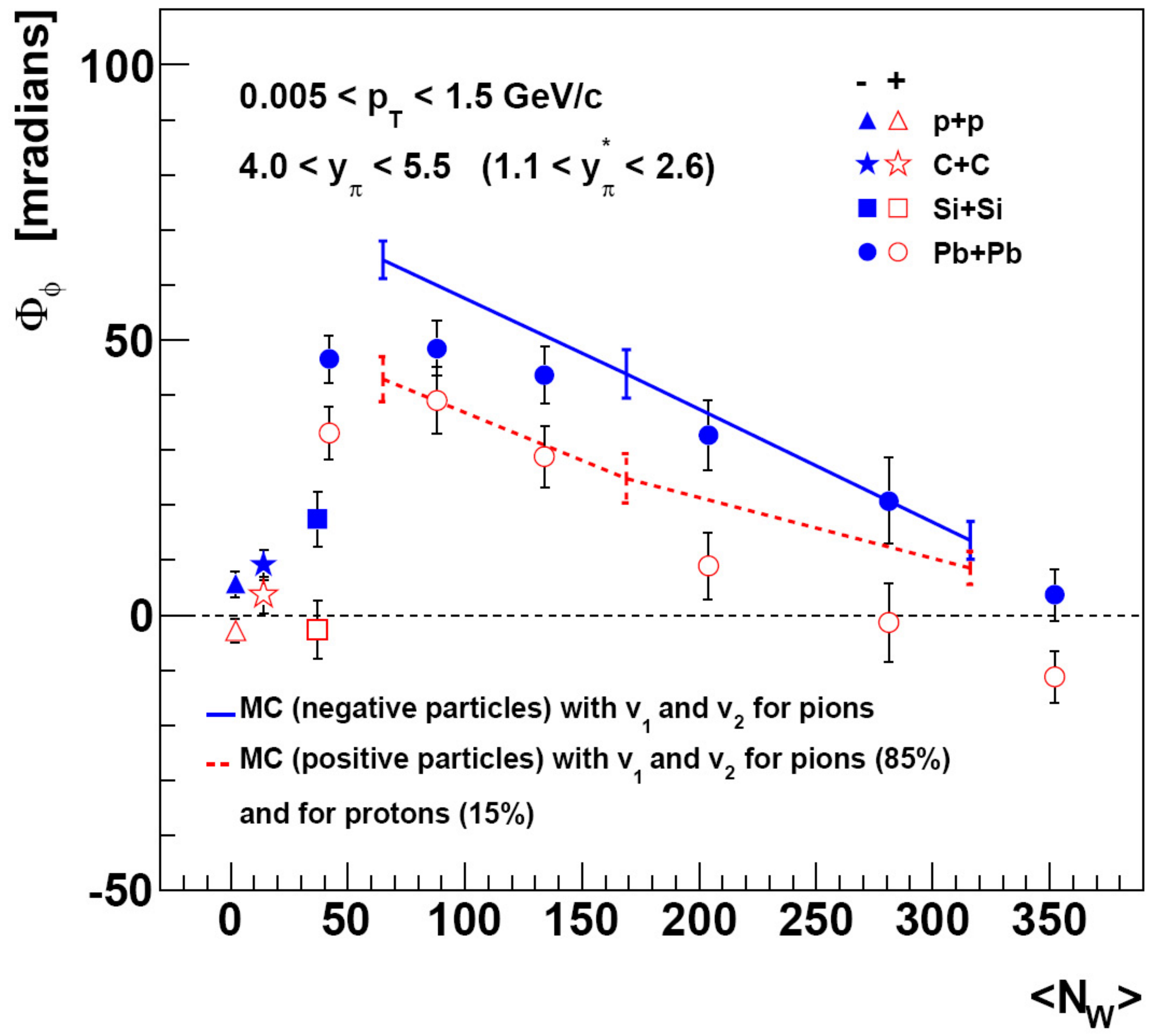}
\caption{$\Phi_\phi$ as a function of the number of wounded nucleons for positively and 
negatively charged particles produced in various colliding systems: p+p, C+C, Si+Si and 
Pb+Pb at 158A GeV. The solid (dashed) line connects  three points which represent the
collective flow effect for negative (positive) particles.}
\label{fig-compare}
\end{figure*}

In our study \cite{Cetner:2010wr} we considered separately the azimuthal correlations caused 
by the collective flow, resonance decays, jets and transverse momentum conservation. In contrast
to all other mechanisms under study, which generate negative values of  $\Phi_\phi$, the 
collective follow produces positive values.  So, it is natural to expect that the correlations
caused by the collective flow are responsible for the experimental signal seen in 
Figs. \ref{fig-data-NW} and \ref{fig-data-energy}. The fact that  $\Phi_\phi$ almost vanishes
for very central Pb+Pb collisions and p+p interactions suggests the same.

The collective flow quantified by the measure $\Phi_\phi$ was studied in 
\cite{Mrowczynski:1999vi}. When the multiplicity distribution is poissonian,
which is true in narrow centrality classes, the measure was found to be 
\be
\label{phi-col-poisson}
\Phi_\phi = \sqrt{ {\pi^2 \over 3} +  \langle N \rangle S }
- {\pi \over \sqrt{3}} ,
\ee
where
\be
S \equiv 2 \Big\langle \sum_{n=1}^{\infty}
\Big({v_n \over n} \Big)^2 \Big\rangle .
\ee
Thus, $\Phi_\phi$ is fully determined by the average Fourier harmonics 
and average particle multiplicity. 

Because of incomplete experimental acceptance in azimuthal angle, we used a simple Monte 
Carlo model instead of the formula (\ref{phi-col-poisson}) to check whether the collective flow 
is indeed responsible for the correlation signal seen in Fig.~\ref{fig-data-NW}. Particle's number 
distribution was poissonian with the average multiplicities of positively and negatively charged 
particles which were measured in a given acceptance window together with $\Phi_\phi$ for 
every centrality. Negatively charged particles were all pions but  among positively charged particles 
there was a 15\% admixture of protons. The estimate was based on predictions of the UrQMD model 
within the NA49 acceptance. The azimuthal angle of each particle was generated from the distribution
\be
\label{v2-dis}
P(\phi) \sim 1 + 2 v_1 {\rm cos}(\phi - \phi_R)  
+2 v_2 {\rm cos}\big(2(\phi - \phi_R) \big) ,
\ee
where $0 \le \phi \le 2\pi$; the reaction plane angle $\phi_R$ of a given event was generated from 
the flat distribution. The Fourier harmonics $v_1$ and $v_2$ in the rapidity domain of interest in 
central, mid-central and peripheral Pb+Pb collisions at 158A GeV had been measured by the NA49 
Collaboration \cite{Alt:2003ab}. The higher Fourier harmonics $v_n$ with $n \ge 3$ were assumed 
to vanish. The predictions of our Monte Carlo model for central, mid-central and peripheral Pb+Pb 
collisions at 158A GeV are compared with the experimental  data on $\Phi_\phi$ in Fig.~\ref{fig-compare}. 

As seen, the  model fairly well estimates the observed $\Phi_\phi$ for both positively and negatively 
charged particles. The main contribution to $\Phi_\phi$ comes from the directed flow represented by 
$v_1$. The difference of $\Phi_\phi$  for positive and negative particles occurs because $v_1$ of 
protons is significantly smaller than that of pions  \cite{Alt:2003ab}. 

\section{Conclusions}

The azimuthal correlations in nucleus-nucleus collisions at CERN SPS, which are quantified 
by the integral measure  $\Phi_\phi$, are strongly dominated by the directed and elliptic flow
generated in the collisions. In the forward rapidity window under study, the directed flow is more
important than the elliptic one.  The difference of $\Phi_\phi$ for positive and for negative particles 
is caused by a 15\% admixture of protons among positive particles. 

\section*{Acknowledgments}

Discussions with Wojciech Broniowski are gratefully acknowledged. This work was partially 
supported by Polish Ministry of Science and Higher  Education under grants N~N202~204638 
and 667/N-CERN/2010/0.

\end{document}